\documentclass[12pt]{article}
\usepackage{amsfonts,amsmath,amsthm, amssymb} 
\usepackage{graphicx}
\usepackage{caption}
\usepackage{setspace}
\usepackage{natbib}
\renewcommand{\qed}{\hfill\square}
\usepackage{color}
\usepackage{floatflt}

\newcommand{\Prob}{\mathbb{P}}
	
\newcommand{\R}{\mathbb{R}}	

\def\O{\mathcal{O}}
\def\o{{\scriptstyle\mathcal{O}}}

\newcommand{\calX}{\mathcal{X}}

\newcommand{\calQ}{\mathcal{Q}}
\newcommand{\calF}{\mathcal{F}}

\def\beqn{\begin{eqnarray*}}
\def\eeqn{\end{eqnarray*}}
\def\beq{\begin{eqnarray}}
\def\eeq{\end{eqnarray}}

\newcommand{\Ex}{\mathbb{E}}

\newcommand{\hatG}{\widehat{G}}
\newcommand{\uc}{\underline{c}}
\newcommand{\oc}{\overline{c}}

\theoremstyle{plain}
\newtheorem{theorem}{Theorem}
\newtheorem{corollary}{Corollary}
\newtheorem{lemma}{Lemma}

\theoremstyle{definition}

\newtheorem{assumption}{Assumption}

\theoremstyle{remark}

\makeatother
\makeatletter
\renewcommand{\@fnsymbol}[1]{\@arabic{#1}}
\makeatother
\title{Asymptotic Distribution and Simultaneous Confidence Bands for Ratios of Quantile Functions}
\author{Fabian Dunker\thanks{School of Mathematics and Statistics, University of Canterbury, Private Bag 4800, Christchurch 8140, New Zealand.}
\and Stephan Klasen\footnote{Department of Economics,  Georg-August-Universit\"at G\"ottingen,
Platz der G\"ottinger Sieben 3, 37073 G\"ottingen, Germany.}
\and Tatyana Krivobokova\footnote{Institute for
  Mathematical Stochastics, Georg-August-Universit\"at-G\"ottingen,
  Goldschmidtstr. 7, 37077 G\"ottingen, Germany. }
}

\date{\today}
\begin{document}

\baselineskip=25pt

\maketitle
\begin{abstract}
\noindent Ratio of medians or other suitable quantiles of two distributions is widely used in medical research to compare treatment and control groups or in economics to compare various economic variables when repeated cross-sectional data are available. Inspired by the so-called growth incidence curves introduced in poverty research, we argue that the ratio of quantile functions is a more appropriate and informative tool to compare two distributions. We present an estimator for the ratio of quantile functions and develop corresponding simultaneous confidence bands, which allow to assess  significance of certain features of the quantile functions ratio. Derived simultaneous confidence bands rely on the asymptotic distribution of the quantile functions ratio and do not require re-sampling techniques. The performance of the simultaneous confidence bands is demonstrated in simulations. Analysis of the expenditure data from Uganda in years 1999, 2002 and 2005 illustrates the relevance of our approach.

\end{abstract} 

\newpage

\doublespacing
\allowdisplaybreaks

\section{Introduction}

Let $X_1$ and $X_2$ be two independent random variables with cumulative distribution functions $F_1$ and $F_2$, respectively. The corresponding quantile functions are given by $Q_j(p)=F_j^{-1}(p)=\inf\{x:F_j(x)\geq p\}$, $j=1,2$. In many applications it is of interest to compare quantiles of two random variables at a given $p\in(0,1)$, which can be done by considering 
$$
g(p)=\frac{Q_2(p)}{Q_1(p)}.
$$
For example, if $X_1$ is income in some population at time $t_1$ and $X_2$ is income at time $t_2>t_1$, then $g(p)$ reports the proportion by which the $p$-quantile of income changed from $t_1$ to $t_2$, with $g(p)>1$ indicating income growth. In medical research one can compare quantiles of some measures obtained in treatment and control groups and then $g(p)$ shows the effect of the treatment on the $p$-quantile. 

In applications $g(p)$ is either considered and interpreted at a fixed $p\in(0,1)$ or the curve $g(p)$, $p\in(0,1)$ is reduced to some number. For example, \cite{ChengWu:2010} as well as \cite{Wu:2010} studied the effect of cancer treatment measured by the ratio of the cancer volumes in the treatment and the control group, the so-called $T/C$-ratio. The $T/C$-ratio can be formed for the mean cancer volume or for a certain quantile of the volume in the treatment and the control group, but typically is not considered as a function of $p$. 
 \cite{DCNZ:05} and \cite{DZ:2005} used the whole curve $g(p)$, $p\in(0,1)$ but only to calculate the mean difference 
$$
\Delta=\Ex(X_1)-\Ex(X_2)=\int_0^1\{Q_1(p)-Q_2(p)\}dp=\int_0^1\left[Q_{1}(p)\left\{1-g(p) \right\}\right]dp
$$
which is known as the average treatment effect (ATE). To obtain $\Delta$,  $\log\{g(p)\}$ is estimated by a smooth function. This approach has been applied to estimate the difference in medical expenditures between persons suffering from diseases attributable to smoking and persons without these diseases.

However, it is clearly more advantageous to view $g(p)$ as a function of $p$. To the best of our knowledge, this has been done only in the poverty research context. In particular, \cite{RavallionChen:03} used the curve 
$$G(p)=\left\{\frac{Q_2(p)}{Q_1(p)}  \right\}^m-1=\{g(p)\}^m-1,\;\;p\in(0,1),\;\;m=\frac{1}{t_2-t_1}\in(0,1]$$
for the analysis of income distributions in developing countries at times $t_1<t_2$ and called $G(p)$ the growth incidence curve (GIC). Poverty reduction can be understood as increasing the incomes of the poor. In this sense poverty is reduced from period $t_1$ to $t_2$, if $G(p)$ takes positive values for all small quantiles up the quantile where the poverty line was located in the first period. Such growth that increases the incomes of poor quantiles has been called ``weak absolute'' pro-poor growth, i.e. growth that is accompanied by absolute poverty reduction without making any statement about the distributional pattern of growth, see \cite{Klasen:08}. 
On the other hand, if $G(p)$ has a negative slope, growth was pro-poor in the relative sense, i.e. the poor benefited (proportionately) more from growth than the non-poor. This means that such growth episodes led to a decrease in inequality and relative poverty. For a detailed discussion of different notions of pro-poor growth we refer to \cite{Ravallion:04} and \cite{Klasen:08}. Growth incidence curves were also applied to non-income data in \cite{GHK:08}. 

Hence, considering the whole curves $g(p)$ or $G(p)$, $p\in(0,1)$ provides more informative comparison of two distributions and can be applied not only in the poverty research context. The goal of this work is to derive the asymptotic distribution of an estimator of $g(p)$ and build simultaneous confidence bands for $g(p)$. Estimation and inference for $G(p)$ is then straightforward.

\cite{DCNZ:05} proposed an estimator for $\log\{g(p)\}$ using smoothing splines. \cite{VDP:2015} extend the work by \cite{DCNZ:05}, employing a Bayesian approach to get a smooth estimator of $h\{g(p)\}$, for some known monotone differentiable function $h$. A much simpler approach, which we pursue, would be to replace the unknown $Q_j(p)$ in $g(p)$ by some estimator $\widehat{Q}_j(p)$, $j=1,2$ to get $\widehat{g}(p)$. There are several quantile estimators available \citep[see e.g.][]{ HD:82,KL:82, Cheng:94}. In this work we employ the classical empirical quantile function.

Apparently, simultaneous inference about the curve $g(p)$, $p\in(0,1)$ is crucial in applications, but has not been considered so far, to the best of our knowledge. \cite{DCNZ:05} rather focused on estimation of the average treatment effect with the help of $\log\{g(p)\}$ and do not discuss inference about $g(p)$. \cite{ChengWu:2010} consider estimation of $g(p)$ at a given $p\in(0,1)$ and build a confidence interval for $g(p)$ using asymptotic normality arguments and several estimators for the variance of $\widehat{g}(p)$.  The Worldbank Poverty Analysis Toolkit (can be found at {http://go.worldbank.org/YF9PVNXJY0}) provides also only point-wise confidence intervals for growth incidence curves, similar in spirit to that of \cite{ChengWu:2010}. More specifically, the confidence statement in this toolkit is constructed for a discretization of $(0,1)$ by $0 < p_1 < p_2 < \ldots < p_k < 1$. For every $p_i$, $i=1,\ldots, k$ expectation and variance for some estimator $\widehat{G}(p_i)$ of $G(p_i)$ are estimated with a bootstrap. Critical values $\uc_i$ and $\oc_i$ are then taken from the corresponding $t$-distribution for some level $\alpha$. This implicitly assumes that $\hatG(p_i)$ is asymptotically normal. The resulting confidence statement has the form
$$
\Prob\{\uc_i \le G(p_i) \le \oc_i\} = 1-\alpha,\;\; \mbox{for each } i=1,2,\ldots,k,
$$
where $\alpha\in(0,1)$ is some pre-specified confidence level. Obviously, these confidence intervals provide inference only at a given $p_i$. 
For example, if we would like to test significance of the poverty reduction (or treatment effect) at the median, it is enough to build a point-wise confidence interval for $G(0.5)=\{g(0.5)\}^m-1$ (or for $g(0.5)$) and check if it includes zero (or one). However, for the test if the growth was pro-poor in the relative sense, a confidence statement about the slope of $G(p)$ has to be made and, hence, simultaneous confidence bands should be considered. That is, the goal is to find such $\uc(p)$ and $\oc(p)$ that
$$
\Prob\left\{\uc(p)\leq G(p)\leq \oc(p)\;\;\mbox{for all}\;\;p\in(0,1)\right\}=1-\alpha.
$$
The difference to the point-wise intervals is that $\uc(p) \le G(p) \le \oc(p)$ holds not only separately for every $p$,  but simultaneously for all $p\in(0,1)$. 

To build simultaneous confidence bands for $g(p)$ or $G(p)$, the analysis of the asymptotic distribution of the function $\widehat{g}(p)$ is necessary. This 
involves the theory of empirical processes which goes back to \cite{Glivenko:33}, \cite{Cantelli:33}, \cite{Donsker:52}, and \cite{KMT:75}. Our analysis builds on results for empirical quantile processes and its simultaneous confidence bands developed in \cite{CR:78}, \cite{CR:84}, and \cite{Csorgo:83}. The main benefit of this approach is that it allows for faster computation of the confidence bands without re-sampling techniques.

The paper is organized as follows. In Section \ref{sec:estimation} we introduce a simple sample counterpart estimator and analyse its asymptotic distribution. This estimator is also used by the World Bank Toolkit. The results about the asymptotic distribution motivates two constructions for asymptotic simultaneous confidence bands presented in Section \ref{sec:conf_bands}. Section \ref{sec:sim} evaluates the small sample properties of our confidence bands by Monte Carlo simulations. Expenditure data from Uganda are analysed with our confidence bands in Section \ref{sec:application} before we conclude in Section \ref{sec:conclusion}.

\section{Estimation and asymptotic distribution}\label{sec:estimation}

Throughout this section we assume that we have i.i.d. samples $X_{1,1}, X_{1,2} \ldots X_{1,n_1}$ of $X_1$ and $X_{2,1}, X_{2,2} \ldots X_{2,n_2}$ of $X_2$. Furthermore, we assume that the samples are stochastically independent of each other. This assumption is justified if the data are collected in two independent groups (e.g. treatment and control) or in repeated cross-sections. Note that there is a related concept of non-anonymous growth incidence curves proposed for panel data in \cite{Grimm:07} and \cite{Bourguignon:11}. Non-anonymous growth incidence curves are built based on two dependent samples and are not treated in this work.

\subsection{Quantile ratio estimator}\label{sec:estimator}
We start by presenting a simple sample  estimator for $g(p)$ and $G(p)$. For $j = 1,2$ we denote the $k$-th order statistic of the sample $X_{j,1}, X_{j,2} \ldots X_{j,n_j}$ by $X_{j,(k)}$. The sample quantile function is the inverse of the right continuous empirical distribution function, which is known to be
\begin{align}\label{eq:sample_quantile_fct}
\widehat{Q}_j(p)= \widehat{F}_j^{-1}(p)=X_{j,(k)}, \;\;\; \mbox{for } \;\; \frac{k-1}{n_j} < p \le \frac{k}{n_j}, \quad k=1, 2, \ldots, n_j,\;\;j=1,2.
\end{align} 
We now define estimators of $g(p)$ and $G(p)$ as
\begin{align}\label{eq:gic_estimator}
\widehat{g}(p)=\frac{\widehat{Q}_2(p)}{\widehat{Q}_1(p)}\;\;\mbox{  and  }\;\;\widehat{G}(y) = \left\{ \widehat{g}(p)\right\}^m-1,\;\;m\in(0,1].
\end{align}

It is well-known that the quantile function and its empirical version are equivariant under strictly monotone transformations. Let us denote by $\calF_j$ and $\calQ_j=\calF^{-1}_j$ the cumulative distribution and quantile functions of $\calX_j=\log(X_j)$, $j=1,2$, respectively. Also, let $\widehat\calQ_j$ be the empirical quantile function of the log-transformed sample $\calX_{j,i}=\log(X_{j,i})$, $i=1,\ldots,n_j$, $j=1,2$. Then, $\calQ_j=\log(Q_j)$, as well as $\widehat{\calQ}_j=\log(\widehat{Q}_j)$, $j=1,2$. 
Consequently, 
\begin{align}
\begin{split}\label{eq:log_trafo}
\log\left\{g(p)\right\}&=\calQ_2(p)-\calQ_1(p),\quad
\log\left\{\widehat{g}(p)\right\}=\widehat\calQ_2(p)-\widehat\calQ_1(p)\\
\log \left\{G(p)+1\right\} &=  m\left\{\calQ_2(p)-\calQ_1(p)\right\},\quad
\log \{\hatG(p) +1 \}=m\{\widehat\calQ_2(p)-\widehat\calQ_1(p)\}.
\end{split}
\end{align}
Hence, a simultaneous confidence band for $g(p)$ can be obtained observing that
$$
\Prob\left\{\uc(p)\leq g(p)\leq \oc(p),\forall p\in(0,1)\right\}=\Prob\left[\log\{\uc(p)\}\leq \calQ_2(p)-\calQ_1(p)\leq \log\{\oc(p)\},\forall p\in(0,1)\right].
$$
Note that the difference of two quantile functions $\Delta(p)=Q_2(p)-Q_1(p)$ is known as quantile treatment effect (QTE), sometimes also named the percentile-specific effect between two populations, see \citet{DZPKC:2006}. To the best of our knowledge, the inference for QTE is usually done at a fixed $p\in(0,1)$, rather than simultaneously.

\subsection{Point-wise asymptotic distribution}\label{sec:point_dist}
We first characterizes the asymptotic distribution of $\hatG(p)$ at a fixed $p\in(0,1)$. The following assumption usually holds for data on income, expenditure, or cancer volume, etc.
\begin{assumption}\label{ass:basic}
Two independent random variables $X_1 > 0\; {a.s.}$ and $X_2>0\; a.s.$ with finite second moments and cumulative distribution functions $F_{1}$ and $F_{2}$ are given together with random samples $X_{j,1}, X_{j,2}, \ldots, X_{j,n_j}$, $j=1,2$. The log-transformed $\calX_j=\log(X_j)$ has the cumulative distribution function $\calF_j$ and density $f_j=\calF_j^{\;'}$, $j=1,2$. The corresponding quantile function $\calQ_j(p)={\calF}^{-1}_j(p)$ has the quantile density $q_j(p)=\calQ^{'}(p)=1/f_j\{\calQ_j(p)\}$, $p\in(0,1)$, $j=1,2$.
\end{assumption}

\begin{theorem}\label{the:ass_lognorm}
Let Assumption \ref{ass:basic} hold and $p\in(0,1)$ be fixed. Moreover, assume $\calF_1$ and $\calF_2$ are continuously differentiable at some $x_1$ and $x_2$, respectively, such that $ \calF_1(x_1)=\calF_2(x_2)=p$ and $f_1(x_1)>0, \;f_2(x_2)>0$. 
\begin{itemize}
\item[(i)] For $\min\{n_1,n_2\} \rightarrow \infty$ the estimator $\hatG(p)+1=\{\widehat{g}(p)\}^m$ is asymptotically log-normal with the parameters $\mu(p)=m\log\{g(p)\}$ and 
\[
\sigma(p)= \sqrt{{m^2p(1-p)}\left[\frac{\{q_1(p)\}^2}{n_1}+\frac{\{q_2(p)\}^2}{n_2}\right]}.
\]
\item[(ii)] If in addition $\calF_1$ and $\calF_2$ are continuously differentiable at some $\tilde{x}_1$ and $\tilde{x}_2$, respectively, such that $\calF_1(\tilde{x}_1)=\calF_2(\tilde{x}_2)=\tilde{p}$, for some $0<p\le \tilde p< 1$, and $f_1(\tilde{x}_1)>0,\;f_2(\tilde{x}_2)>0$, then the asymptotic distribution of $\{\hatG(p)+1,\hatG(\tilde p)+1\}$ is bivariate log-normal with the parameters $\{\mu(p),\mu(\tilde p)\}$ and  
$$
\sigma(p,\tilde{p})={m^2p(1-\tilde{p})}\left\{\frac{q_1(p)q_1(\tilde{p})}{n_1}+\frac{q_2(p)q_2(\tilde{p})}{n_2}\right\}.
$$
\end{itemize}
\end{theorem}

\begin{corollary}\label{cor:ass_norm}
Under the assumptions of Theorem \ref{the:ass_lognorm} we have asymptotic normality for $\hatG(p)+1=\{\widehat{g}(p)\}^m$ in the sense that 
$$
\frac{\hatG(p)+1-\{g(p)\}^m}{\{g(p)\}^{m} \sigma(p)}\stackrel{\mathcal{D}}{\longrightarrow} \mathcal{N}(0,1) 
$$
converges in distribution to a standard normal random variable for $\min\{n_1,n_2\} \rightarrow \infty$ and for any fixed $p\in(0,1)$.
\end{corollary}

The World Bank Toolkit and \citet{ChengWu:2010} implicitly employ the asymptotic normality of $\widehat{G}(p)$ and $\widehat{g}(p)$ to build point-wise confidence intervals, but use different variance estimators, based either on bootstrap or on certain approximations. 
To the best of our knowledge, the result of Corollary \ref{cor:ass_norm} is new. Note also that $\sigma(p)$ depends on unknown $q_j(p)$, $j=1,2$, which have to be consistently estimated in practice.

Theorem \ref{the:ass_lognorm} and Corollary \ref{cor:ass_norm} provide two different ways for deriving point-wise confidence statements about $G(p)$ (or about $g(p)$ by setting $m=1$). We can approximate the distribution of $\hatG(p)+1=\{\widehat{g}(p)\}^m$ for a fixed $p\in(0,1)$ either by a  log-normal or by a normal distribution. 
However, the log-normal approximation is preferable for positive random variables. Indeed, $X_j>0 \;a.s.$, $j=1,2$ implies 
$g(p) \in [0,\infty)$ for all $p\in (0,1)$. Hence, a normal approximation of the distribution of $\widehat{G}(p)+1=\{\widehat{g}(p)\}^m$ puts probability mass outside of $[0,\infty)$. This can cause confidence intervals to take impossible values, in particular in small samples, and affect the actual coverage of the band. Taking a log-normal approximation helps to avoid this. We use the log-normal approximation implicitly in our constructions of simultaneous confidence bands in Section \ref{sec:conf_bands}.

\subsection{Approximation by Brownian bridges}\label{sec:uni_dist}

In the previous Section \ref{sec:point_dist} derivation of the confidence statements about $G(p)$ or $g(p)$ at one or at a finite number of points reduces to finding the limiting distribution of $\widehat\calQ_2(p)-\widehat\calQ_1(p)$ at a fixed $p\in(0,1)$. To obtain confidence statements about $G(p)$ or $g(p)$ that hold for all $p\in(0,1)$ simultaneously, we need to find the limiting distribution of $\widehat\calQ_2(p)-\widehat\calQ_1(p)$, which is treated as a stochastic process indexed in $p \in (0,1)$.

Let us define the following stochastic process
\begin{align*}
D_{n_1,n_2}(p;s)= \sqrt{\frac{n_1 n_2}{n_1 + s^2 n_2}} \left\{s\;\frac{\widehat{\calQ}_1(p)-\calQ_1(p)}{q_1(p)}-\frac{\widehat{\calQ}_2(p)-\calQ_2(p)}{q_2(p)}
\right\},\;\;p\in(0,1),
\end{align*}
where $s>0$ is a fixed scaling parameter independent of $n$ needed later for technical reasons. For the analysis of this process we need the following set of assumptions on $X_1$ and $X_2$.
\begin{assumption}\label{ass:diff}
The cumulative distribution functions $\calF_j$ of the log-transformed $\calX_j=\log(X_j)$, $j=1,2$ are twice differentiable on $(a,b)$, where $a = \sup\{x:\calF_j(x) = 0\}$, $b = \inf\{x:\calF_j(x) = 1\}$, $-\infty\leq a<b\leq\infty$ and $f_j > 0$ on $(a,b)$. In addition, there exists some $0 < \gamma < \infty$ such that
\begin{align}\label{eq:bound_quant_dens}
\sup_{x \in (a,b)} \calF_{j}(x)\{1-\calF_{j}(x)\}\left|\frac{f^{\;'}_{j}(x)}{\left\{f_j(x)\right\}^2}\right| \le \gamma,\;\;j=1,2.
\end{align}
\end{assumption}

\begin{assumption}\label{ass:tail}
For
$
A_j=\limsup_{x\searrow a} f_{j}(x) \le \infty$ and $B_j=\limsup_{x\nearrow b} f_{j}(x) \le \infty
$, $j=1,2$
one of the following conditions hold
\begin{itemize}
\item[(i)] $\min(A_j,B_j)>0$
\item [(ii)] If $A_j=B_j=0$, then $f_{j}$ is non-decreasing on an interval to the right of $a$ and non-increasing on an interval to the left of $b$.
\end{itemize}
\end{assumption}

If $X_1$ and $X_2$ are log-normal, as typically the case for income, expenditure and similar positive random variables, then $f_j$ is the density of a normal distribution. Hence, existence, positivity and differentiability of $f_j$ on $\R$ are trivially fulfilled. The supremum in \eqref{eq:bound_quant_dens} is $1$ for normally distributed random variables independent of expectation and variance. The property in Assumption \ref{ass:tail} 
is called tail-monotonicity. 
For normal distributions $A_j=B_j=0$ and Assumption \ref{ass:tail} (ii) obviously holds. 

The following result shows that $D_{n_1,n_2}(p;s)$ converges uniformly to a Brownian bridge $B(p)$.
Recall that a Brownian bridge is a standard Wiener process $W(p)$ with $W(0) = W(1) = 0$, i.e. $B(p)=W(p)-pW(1)$, $p\in[0,1]$. In particular, $B(p) \sim \mathcal{N}(0,p-p^2)$ and $Cov\{B(p), B({\tilde p})\} = p(1-\tilde p)$ for all $0 \le p \le \tilde p \le 1$. 

\begin{theorem}\label{the:gic_process}
Let Assumptions \ref{ass:basic} and \ref{ass:diff} hold and set $n= \min\{n_1,n_2\}$. Then a series of Brownian bridges $B_{n_1,n_2}$ can be defined such that for any fixed $s$
\begin{align*}
\sup_{p \in [\delta_n,1-\delta_n]} \Big| D_{n_1,n_2}(p;s)- B_{n_1,n_2}(p)\Big| \stackrel{a.s.}{=} \O\left\{n^{-1/2}\log(n)\right\}
\end{align*}
with $\delta_n = 25\;n^{-1}\log\log(n)$. If in addition Assumption \ref{ass:tail} holds, a Brownian bridge $B_{n_1,n_2}$ can be defined such that in case of Assumption \ref{ass:tail} (i)
\begin{align*}
\sup_{p \in [0,1]} \Big|D_{n_1,n_2}(p;s) - B_{n_1,n_2}(p)\Big| \stackrel{a.s.}{=} \O\left\{n^{-1/2}\log(n)\right\}
\end{align*}
and in case of Assumption \ref{ass:tail} (ii)
\begin{align*}
\sup_{p \in (0,1)} \Big| D_{n_1,n_2}(p;s) - B_{n_1,n_2}(p)\Big|\stackrel{a.s.}{=}
\begin{cases}
\O\left\{n^{-1/2}\log(n)\right\}& \mbox{if } \gamma < 2\\
\O\left[n^{-1/2}\{\log\log(n)\}^\gamma \{\log(n)\}^{(1+\varepsilon)(\gamma-1)}\right] & \mbox{if } \gamma \ge 2
\end{cases}
\end{align*}
for arbitrary $\varepsilon > 0$.
\end{theorem}

For example, if $X_j$ are approximately log-normal in a way that $\log(X_j)$ has the tail behavior of a normal variable, then according to Theorem \ref{the:gic_process} the process $D_{n_1,n_2}(p;s)$ converges to a Brownian bridge simultaneously on $(0,1)$ with the rate $O\{n^{-1/2} \log(n)\}$.

Constructing confidence sets for $g(p)$ or $G(p)=\{g(p)\}^m-1$ requires knowledge of the asymptotic distribution of $\widehat\calQ_1(p)-\widehat\calQ_2(p)=\log\{\widehat{g}(p)\}=m^{-1}\log\{\widehat{G}(p)+1\}$, while $D_{n_1,n_2}(p;s)$ in Theorem \ref{the:gic_process} contains $s\widehat\calQ_1(p)/q_1(p)-\widehat\calQ_2(p)/q_2(p)$ instead. Therefore, let us  consider 
$$
D^*_{n_1,n_2}(p;s)= 2\sqrt{\frac{n_1 n_2}{n_1 + s^2 n_2}} \;\frac{\widehat{\calQ}_1(p)-\calQ_1(p)-\left\{\widehat{\calQ}_2(p)-\calQ_2(p)\right\}}{q_1(p)/s+q_2(p)}.
$$
and discuss the choice of $s$. First, introduce the following assumption.

\begin{assumption}\label{ass:scale}
There exists a constant $s>0$ such that the quantile densities satisfy $q_1(p)=sq_2(p)$, $p\in(0,1)$.
\end{assumption}
Obviously, under Assumption \ref{ass:scale} we have that
$$
D^*_{n_1,n_2}(p;s)=D_{n_1,n_2}(p;s)=\sqrt{\frac{n_1 n_2}{n_1 +  s^2n_2}} \;\frac{\widehat{\calQ}_1(p)-\calQ_1(p)-\left\{\widehat{\calQ}_2(p)-\calQ_2(p)\right\}}{q_2(p)}
$$ 
and Theorem \ref{the:gic_process} can be applied to get the asymptotic distribution of $\widehat\calQ_1(p)-\widehat\calQ_2(p)$ and hence the simultaneous confidence bands for $G(p)$ or $g(p)$. 

It is shown in the Appendix, that if Assumption \ref{ass:scale} is true, then
\beq
\label{eq:s}
s=\frac{\int_{-\infty}^\infty \{f_2(x)\}^2dx}{\int_{-\infty}^\infty \{f_1(x)\}^2dx}.
\eeq

Moreover, if the $\calX_j$ have distribution from the location-scale family of distributions with locations $\mu_j$ and scales $\sigma_j<\infty$, $j=1,2$, then Assumption \ref{ass:scale} implies that $s\propto \sigma_1/\sigma_2$. This can be seen directly from (\ref{eq:s}) applying the change of variable $y=\mu_j+\sigma_jx$. Also, let $\widetilde\calQ_j$ denote the quantile function of 
$
\{\calX_{j}-\mu_j\}/{\sigma_j}
$
and $\tilde{q}_j$ the corresponding quantile density. 
Then, $\calQ_j(p)=\mu_j+\sigma_j\widetilde\calQ_j(p)$ and therefore 
$q_j(p)=\sigma_j\;\tilde{q}_j(p)$, $p\in(0,1)$, $j=1,2$. In particular, Assumption \ref{ass:scale} implies that $\tilde{q}_1\propto \tilde{q}_2$ and thus the distributions of $\calX_1$ and $\calX_2$ differ only in location and scale parameters.

For example, if $X_j$ are both log-normally distributed with arbitrary location parameters and scale parameters $\sigma_j$, then $\log(X_j)=\calX_j$, $j=1,2$ are normally distributed and $s=\sigma_1/\sigma_2$. In applications, to check if distributions of $\calX_1$ and $\calX_2$ differ only in the location and scale, one can inspect the QQ-plot of standardised log-transformed data.

If the quantile densities are not proportional, that is, Assumption \ref{ass:scale} is not fulfilled, we have to handle the term
$$
D_{n_1,n_2}^*(p;s)-D_{n_1,n_2}(p;s)= \frac{{q}_1(p)-s\;{q}_2(p)}{{q}_1(p)+s\;{q}_2(p)}\;\sqrt{\frac{n_1 n_2}{n_1 +  s^2n_2}}\;\left\{\frac{\widehat{\calQ}_1(p)-\calQ_1(p)}{q_1(p)/s}+\frac{\widehat{\calQ}_2(p)-\calQ_2(p)}{q_2(p)}
\right\}.
$$

\begin{lemma}\label{lem:itereated_log}
Under  Assumptions \ref{ass:basic}, \ref{ass:diff} and \ref{ass:tail} 
\begin{align*}
\limsup_{n_1,n_2 \rightarrow \infty}\left(\log \log\sqrt{\frac{n_1 n_2}{n_1 +  s^2n_2}} \right)^{-1/2}& \sup_{p\in(1/n,1-1/n)}\left|D_{n_1,n_2}^*(p;s)-D_{n_1,n_2}(p;s)\right|\\
&\stackrel{a.s.}{\le} \frac{4^\nu}{\sqrt{2}}\sup_{p\in(1/n,1-1/n)}\left|  \frac{{q}_1(p)-s\;{q}_2(p)}{{q}_1(p)+s\;{q}_2(p)} \{p(1-p)\}^\nu\right|
\end{align*}
for all $\nu \in [0,1/2)$.
\end{lemma}

Note that the bound on the right hand side is always smaller are equal $1/\sqrt{2}$ for every $\nu \in [0,1/2)$. Since $q_1$ and $q_2$ are usually similar functions in applications, much smaller bounds can be expected.

\section{Simultaneous confidence bands}\label{sec:conf_bands}

Based on the results of the previous section, we can derive simultaneous confidence bands for $\calQ_2(p)-\calQ_1(p)=\log\{g(p)\}=m^{-1}\log\{G(p)+1\}$ and transform them into simultaneous confidence bands for $g(p)$ or $G(p)$. Note that simultaneous confidence bands for the quantile treatment effect $Q_2(p)-Q_1(p)$ follow immediately. We make use of Theorem \ref{the:gic_process} and Lemma \ref{lem:itereated_log} from the last section, as well as the Kolmogorov distribution
\begin{align}\label{eq:conf_bb}
\Prob\left(\sup_{p \in [0,1]}\left|B(p)\right|\le c\right) = \sum_{k=-\infty}^\infty (-1)^k e^{-2k^2c^2}. 
\end{align}
Throughout this section we assume a confidence level $\alpha$ and denote the corresponding critical value for the Brownian bridge by $c_\alpha$ such that $\Prob\left(\sup_{p \in [0,1]}\left|B(p)\right|\le c_\alpha\right) = 1-\alpha$. In addition, we denote by $c_s$ an asymptotically almost sure upper bound  from Lemma \ref{lem:itereated_log}
\[
c_s = \inf_{0\le\nu\le 1/2-\delta} \left(\log \log\sqrt{\frac{n_1 n_2}{n_1 +  s^2n_2}} \right)^{1/2} \frac{4^\nu}{\sqrt{2}}\sup_{p\in(1/n,1-1/n)}\left|  \frac{{q}_1(p)-s\;{q}_2(p)}{{q}_1(p)+s\;{q}_2(p)} [p(1-p)]^\nu\right|.
\]
with some $\delta > 0$.

In the following, we present two ways of using the approximation by Brownian bridges for the construction of simultaneous confidence band for $\calQ_2(p)-\calQ_1(p)$. Similar approaches for the quantile function have been explored in \citet{CR:84}. 

\subsection{Confidence bands with quantile density estimation}

The first approach to the construction of confidence bands relies on the following argument
\begin{align*}
1-\alpha &\approx \Prob\left(\left|D_{n_1,n_2}(p;s)\right|\le c_\alpha,\; \mbox{ for all  } 0<p<1\right)\\
&\le \Prob\left(\left|D^*_{n_1,n_2}(p;s)\right|\le c_\alpha + c_s,\;\mbox{ for all  } 0<p<1\right)\\
&= \Prob\Bigg[\left|\calQ_2(p)-\calQ_1(p)-\left\{\widehat{\calQ}_2(p)-\widehat\calQ_1(p)\right\}\right|\\
&\hspace{100pt}\le  (c_\alpha +c_s)\sqrt{\frac{n_1 + s^2 n_2}{n_1 n_2}}\;\;\frac{q_1(p)/s+q_2(p)}{2},\;\mbox{ for all  } 0<p<1 \Bigg].
\end{align*}
The quantities $q_j(p)$, $j=1,2$ are unknown and have to be estimated. Various nonparametric methods for the estimation of $q_j(p)$  have been proposed, typically based on kernel density estimation, see e.g. \cite{CMHD:1991}, \cite{Jones:92}, \cite{Cheng:95}, \cite{ChengParzen:97}, \cite{Soni:2012}, and \cite{CDD:2016}. We make the following assumption on the densities.

\begin{assumption}\label{ass:order} 
The densities  $f_j$, $j=1,2$ fulfill
$$
\sup_{x\in(a,b)}\frac{\left[\calF_j(x)\{1-\calF_j(x)\}\right]^2}{f_j(x)}  < \infty \qquad \mbox{and} \qquad \sup_{x\in(a,b)}\left|f_j^{''}(x)\right|< \infty.
$$
\end{assumption}

Now we can get the simultaneous confidence bands for the difference of two quantile functions.
\begin{theorem}\label{the:gic_conf1}
Let Assumptions \ref{ass:basic}, \ref{ass:diff}, \ref{ass:tail} and \ref{ass:order} hold and let $K$ be a second order kernel with support in $[-1/2,1/2]$. For $j = 1,2$ set
\[
\widehat{q}_j(p)= h_{n_j}^{-1}\int_0^1 K\left(\frac{y-z}{h_{n_j}}\right) d\widehat\calQ_j(z).
\]
Then a series of Brownian bridges $B_{n_1,n_2}$ can be defined such that for any fixed $s$
$$
\sup_{p\in[\varepsilon_n,1-\varepsilon_n]}\left|\sqrt{\frac{n_1 n_2}{n_1 + s^2 n_2}} \left\{\frac{\widehat{\calQ}_1(p)-\calQ_1(p)}{\widehat{q}_1(p)/s}-\frac{\widehat{\calQ}_2(p)-\calQ_2(p)}{\widehat{q}_2(p)}\right\}-B_{n_1,n_2}(p)\right|\overset{a.s.}{=}\o\left\{\frac{\sqrt{\log\log(n)}}{n^\delta}\right\}
$$
and for 
\beq
\label{eq:calpha}
{c}_\alpha^*(p)=(c_\alpha +{c}_{s})\sqrt{\frac{n_1 + s^2 n_2}{n_1 n_2}}\;\;\frac{\widehat{q}_1(p)/s+\widehat{q}_2(p)}{2}
\eeq
we get  
\begin{align}
\label{eq:gic_conf1}
1-\alpha\leq&\lim_{n_1,n_2\rightarrow\infty} \Prob\left\{\widehat{\calQ}_2(p)-\widehat\calQ_1(p)-c_\alpha^*(p)\leq \calQ_2(p)-\calQ_1(p)\right.\\
&\hspace{100pt}\left.\le\widehat{\calQ}_2(p)-\widehat\calQ_1(p)+ c_\alpha^*(p),\;p\in(\varepsilon_n, 1-\varepsilon_n) \right\}\nonumber
\end{align}
with $h_{n_j} = n_j^{-\eta}$, $n = \min\{n_1,n_2\}$, $\varepsilon_n = n^{-\beta}$, $3\beta +\delta < \eta<1/2$, and $\eta/2 + \delta + 2 \beta < 1/2$.
\end{theorem}

Note that if Assumption \ref{ass:scale} holds, then ${c}_{s}$ in (\ref{eq:calpha}) is set to zero and $s$ is chosen as in \eqref{eq:s}. Simultaneous confidence bands (\ref{eq:gic_conf1}) are given for the difference of two quantile functions, known as the quantile treatment effect. To get simultaneous confidence bands for $g(p)$ and $G(p)$ recall that $\calQ_2(p)-\calQ_1(p)=\log\{g(p)\}=m^{-1}\log\{G(p)+1\}$ so that
\beqn
&&\Prob\left\{\widehat{\calQ}_2(p)-\widehat\calQ_1(p)-c_\alpha^*(p)\leq \calQ_2(p)-\calQ_1(p)\le\widehat{\calQ}_2(p)-\widehat\calQ_1(p)+ c_\alpha^*(p),\;p\in(\varepsilon_n, 1-\varepsilon_n) \right\}\\
&=& \Prob\left\{\exp(-c_\alpha^*(p))\widehat{g}(p)\leq g(p)\leq \exp(c_\alpha^*(p))\widehat{g}(p),\;p\in(\varepsilon_n, 1-\varepsilon_n) \right\}\\
&=& \Prob\left[\left\{\widehat{G}(p)+1\right\}\exp(-c_\alpha^*(p)m)-1\leq G(p)\leq \left\{\widehat{G}(p)+1\right\}\exp(c_\alpha^*(p)m)-1,\;p\in(\varepsilon_n, 1-\varepsilon_n)\right].
\eeqn

\subsection{Direct confidence bands}

The confidence band above depends on nonparametric estimation of quantile densities. Two smoothing parameters $h_{n_j}$, $j=1,2$ have to be chosen, which might be unfavourable in applications. This can be avoided with the alternative construction of confidence bands given in the following theorem.

\begin{theorem}\label{the:gic_conf2}
Let Assumption \ref{ass:basic} and \ref{ass:diff} hold. Then
\begin{align*}
1-\alpha = \lim_{n_1,n_2 \rightarrow \infty} \Prob\Bigg\{\widehat{\calQ}_2&\left(p-\frac{c_\alpha}{\sqrt{2n_2}}\right)- \widehat{\calQ}_1\left(p+\frac{c_\alpha}{\sqrt{2n_1}}\right) \leq\calQ_2(p)-\calQ_1(p)\\
&\leq \widehat{\calQ}_2\left(p+\frac{c_\alpha}{\sqrt{2n_2}}\right)- \widehat{\calQ}_1\left(p-\frac{c_\alpha}{\sqrt{2n_1}}\right);\;\varepsilon_n \le y \le 1-\varepsilon_n\Bigg\},
\end{align*}
with $\varepsilon_n = n^{-1/2+\delta}$ for any $\delta \in (0,1/2)$.
\end{theorem}
Theorem \ref{the:gic_conf2} requires fewer assumptions than Theorem \ref{the:gic_conf1}, but there is no explicit convergence rate given. However, these confidence bands give good results in numerical simulations. To obtain simultaneous confidence bands for $g(p)$ or $G(p)$ use $\calQ_2(p)-\calQ_1(p)=\log\{g(p)\}=m^{-1}\log\{G(p)+1\}$.

\section{Simulation study}\label{sec:sim}
We evaluate the properties of the confidence bands by using synthetic data and building confidence bands for growth incidence curves $G(p)$. Confidence bands for the quantile treatment effect and $g(p)$ are equivalent. We consider two settings and in both of them fix $m=1$. In the first setting $X_1$ and $X_2$ are drawn from log-normal distributions.
Thereby, $X_{1}$ has location parameter $0$ and scale parameter $\sigma_1 = 0.7$, while $X_{2}$ has location parameter $0.8$ and scale parameter $\sigma_2 =1$.
As already discussed, Assumption \ref{ass:scale} holds in this example with $s=\sigma_1/\sigma_2=0.7$. This value is estimated in the simulations, while $c_s$ is set to zero. In the second setting, $X_1$ is as in the first setting, while  $X_2$ is drawn from the gamma distribution with the shape parameter $2$ and scale parameter $1$. In this setting Assumption \ref{ass:scale} does not hold and $c_s$ is estimated for the plug-in confidence bands.

We considered four sample sizes $n_1=n_2=n\in\{100,1\,000,5\,000,10\,000\}$. For probability values $p\in(0,1)$ we used an equidistant grid of length $100$ to build the confidence bands; setting the grid length to $n$ does not change the results significantly, but increases the computation time in Monte Carlo simulations. The results are based on the Monte Carlo samples of size $5\,000$. The following Table \ref{table1} summarizes the actual coverage probability with simulated data for $1-\alpha=0.95$. The results are given in both settings for the confidence bands with plug-in estimators, for the direct confidence bands and for the confidence bands built with the World Bank algorithm.\\

\begin{table}[h!]
\begin{center}
\begin{tabular}{|c||c|c|c||c|c|c|}
\hline
&\multicolumn{3}{l|} {Setting 1}&\multicolumn{3}{l|} {Setting 2}\\\hline
Sample size $n$ & Plug-in &  Direct  &World Bank& Plug-in &  Direct  &World Bank\\ \hline
$100$ & 0.888 & 0.965& 0.460& 0.893 & 0.964& 0.386\\ \hline
$1\,000$ & 0.975  & 0.960& 0.286& 0.958  & 0.960& 0.177\\ \hline
$5\,000$  & 0.980  & 0.959& 0.343& 0.969  & 0.960& 0.267 \\\hline
$10\,000$ & 0.984  & 0.960& 0.425& 0.973  & 0.964& 0.390 \\\hline
\end{tabular}
\caption{Coverage probability of the plug-in, direct and World Bank confidence bands.}
\label{table1}
\end{center}
\end{table}

First of all, the coverage of the confidence bands obtained with the World Bank algorithm is way too small. The reason is that we tested simultaneous coverage, while the World Bank algorithm constructs only point-wise confidence bands.

The actual coverage probability of all our constructions (about $0.96$) is slightly larger than the theoretical probability $0.95$, except for the plug-in confidence bands for $n=100$, where the coverage is lower than the nominal. This can be attributed to the quality of the nonparametric estimates of the quantile densities in small samples, as also expected from Theorem \ref{the:gic_conf1}. Once the sample size is large, both confidence bands perform very similar, even with the estimated correction $c_s$ for the plug-in bands in the second setting.

\begin{figure}[h!]
\includegraphics[width=0.49\textwidth]{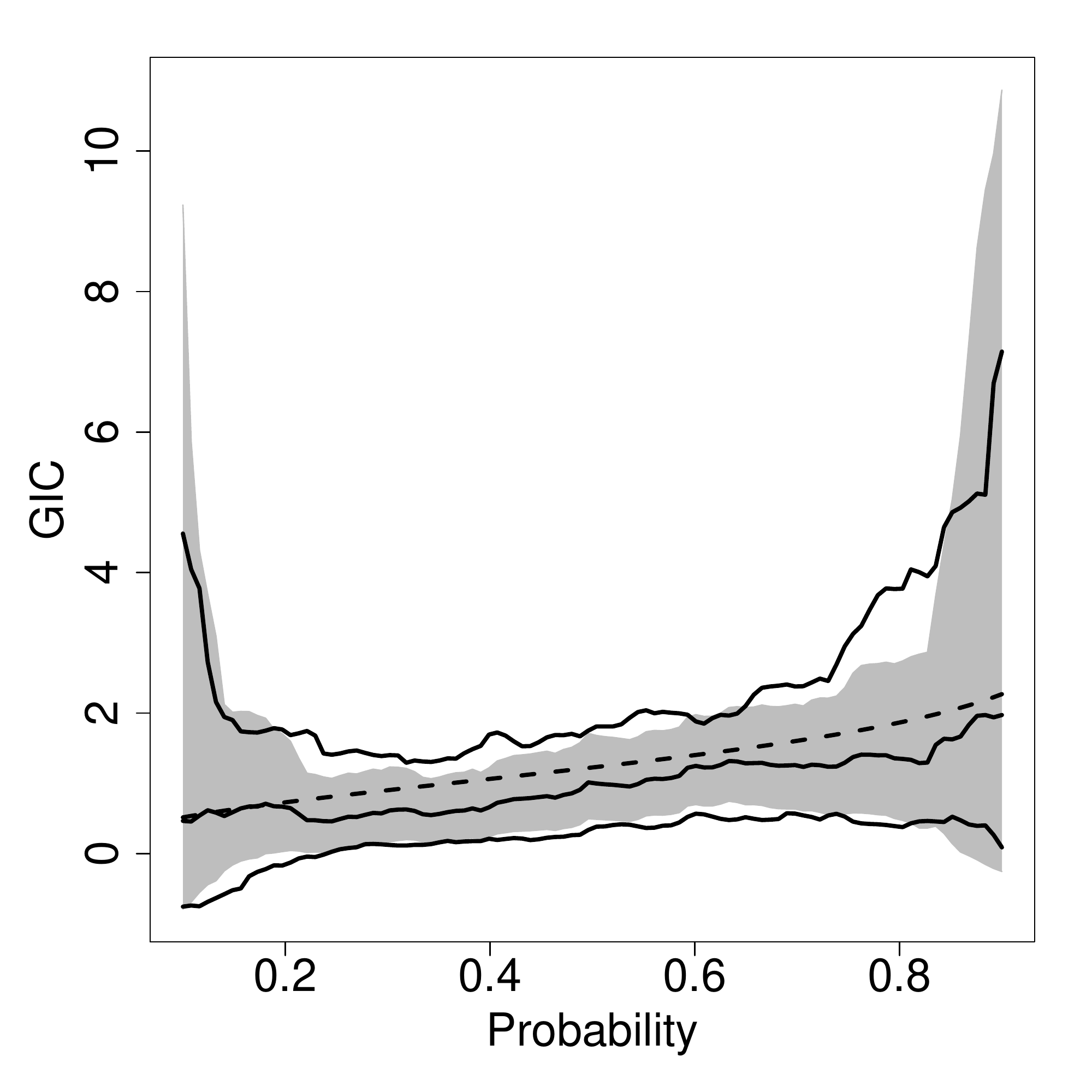}
\quad
\includegraphics[width=0.49\textwidth]{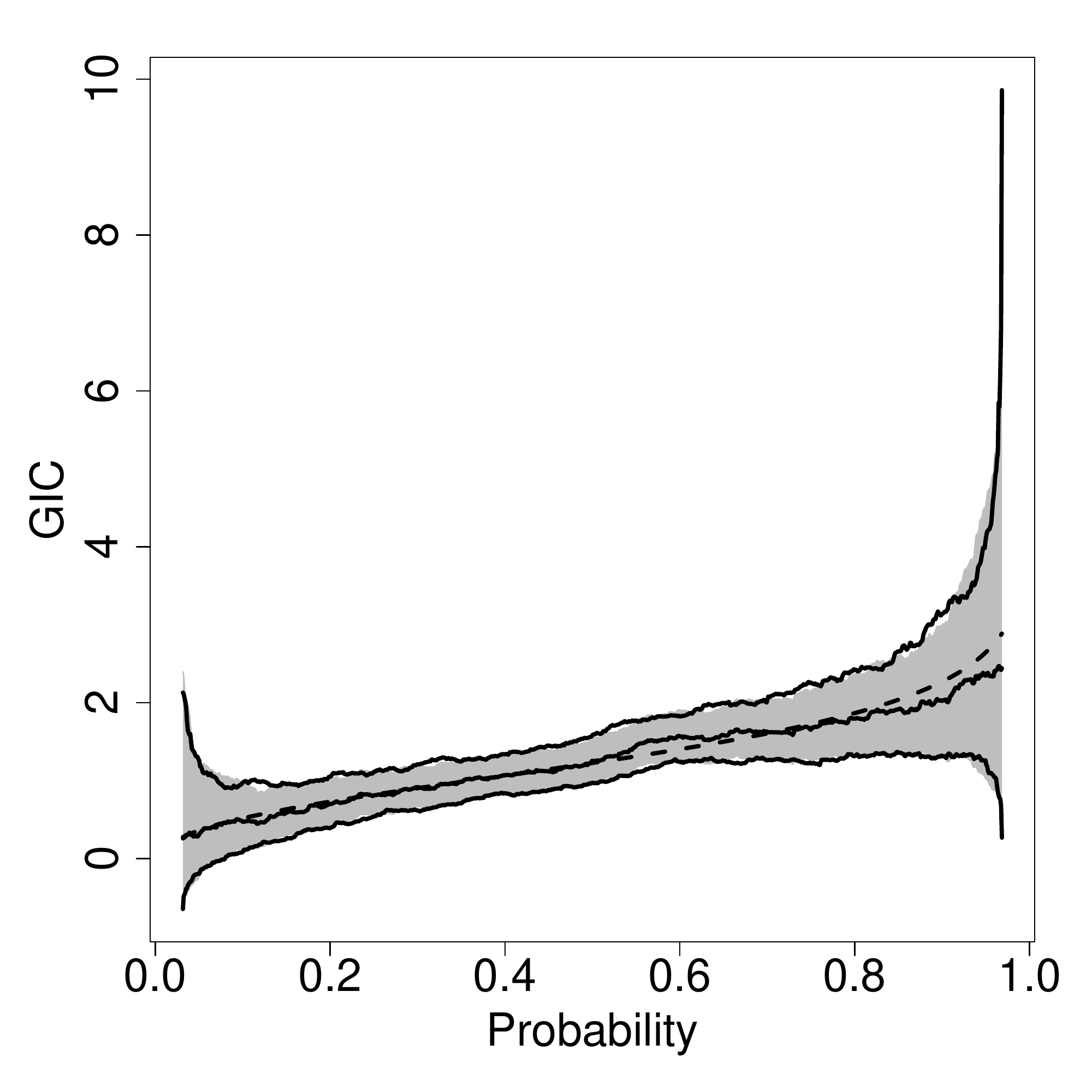}
\caption{Estimates for growth incidence curves and $95\%$ simultaneous confidence bands for  $n = 100$ (left) and $n=1000$ (right). Each plots shows the ture growth incidence curve (dahsed), its estimator (bold), plug-in confidence bands (grey area) and direct confidence bands (bold).}\label{fig:plots_uniform}
\end{figure}

The plots in Figure \ref{fig:plots_uniform} show typical estimates from the first setting together with $95\%$ plug-in and direct confidence bands for $n=100$ (left) and $n=1\,000$ (right). The true growth incidence curve $G(p)$ is the dashed line, while its estimate is the solid line. Plug-in confidence bands are shown as a grey area, while direct confidence bands are solid lines enveloping the growth incidence curve. In accordance with the simulation results, plug-in confidence bands are somewhat narrower for small $n=100$, while for $n=1\,000$ both confidence bands are nearly indistinguishable. As stated in Theorem \ref{the:gic_conf1} and Theorem \ref{the:gic_conf2} the confidence bands are not defined for $p$ close to $0$ and close to $1$. The plots show the bands for probabilities $p$ between $\varepsilon_n$ and $1-\varepsilon_n$.

\section{Application to household data}\label{sec:application}

Our work is motivated by the application of growth incidence curves to the evaluation of pro-poorness of growth in developing countries. Absolute poverty is reduced if the growth incidence curve $G(p)$ is positive for all income quantiles below the poverty line and such growth is called pro-poor using the weak absolute definition mentioned in the introduction. In this case, there is some income growth for the poor and absolute poverty is reduced. In addition, relative poverty is reduced if $G(p)$ has a negative slope, such growth is called pro-poor using the relative definition as it is associated with declining inequality and declining relative poverty.

We analyse data from the Uganda National Household Survey for the years $1992$, $2002$, and $2005$. This is a standard multi-purpose household survey that is regularly conducted to monitor trends in poverty and inequality and its most important correlates. The sample sizes are $n_{1992} = 9923$, $n_{2002} = 9710$, and $n_{2005} = 7421$. We measure welfare by household expenditure per adult equivalent in $2005/2006$ prices and compute the related growth incidence curves.

\begin{figure}[h!]
\includegraphics[scale=0.255]{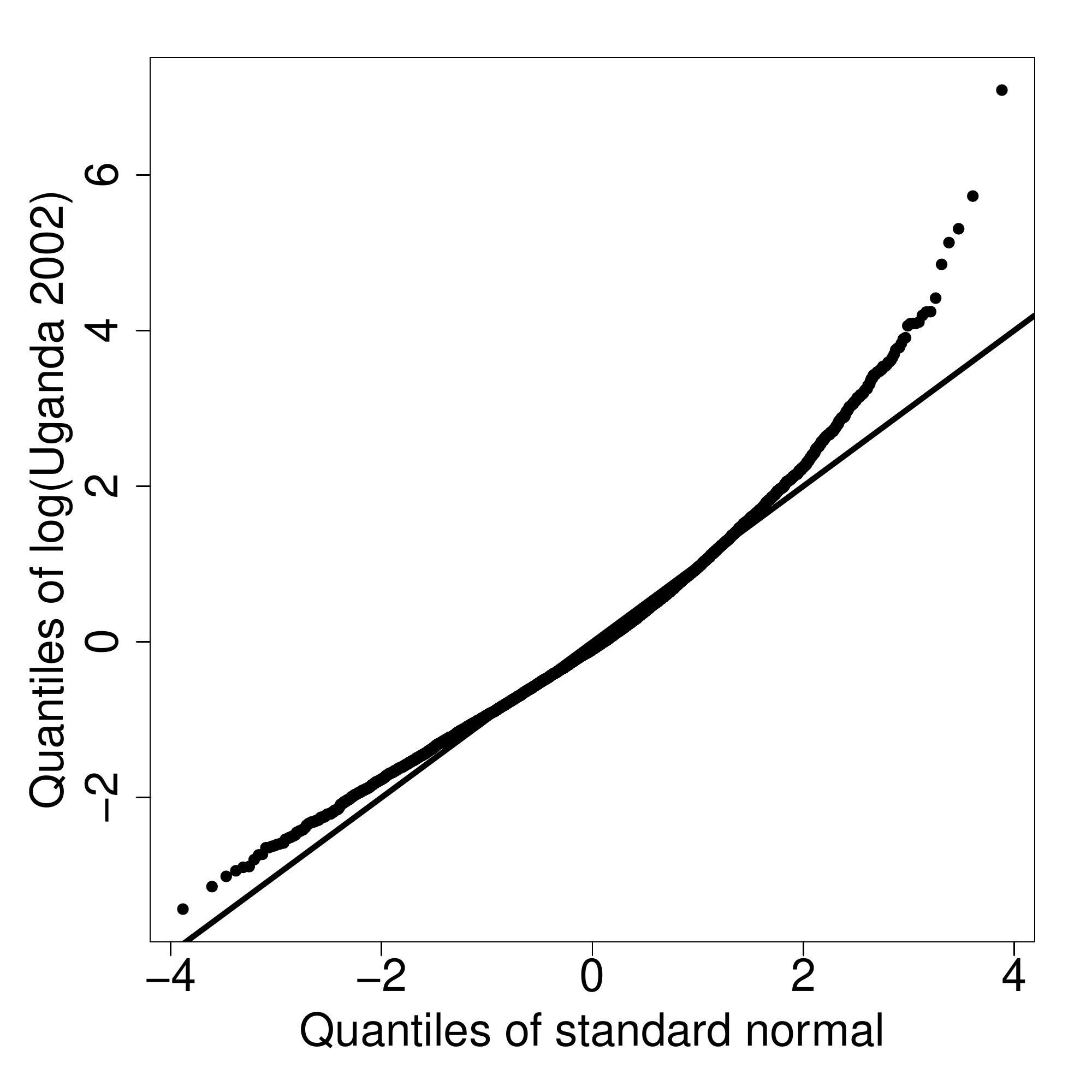}
\includegraphics[scale=0.255]{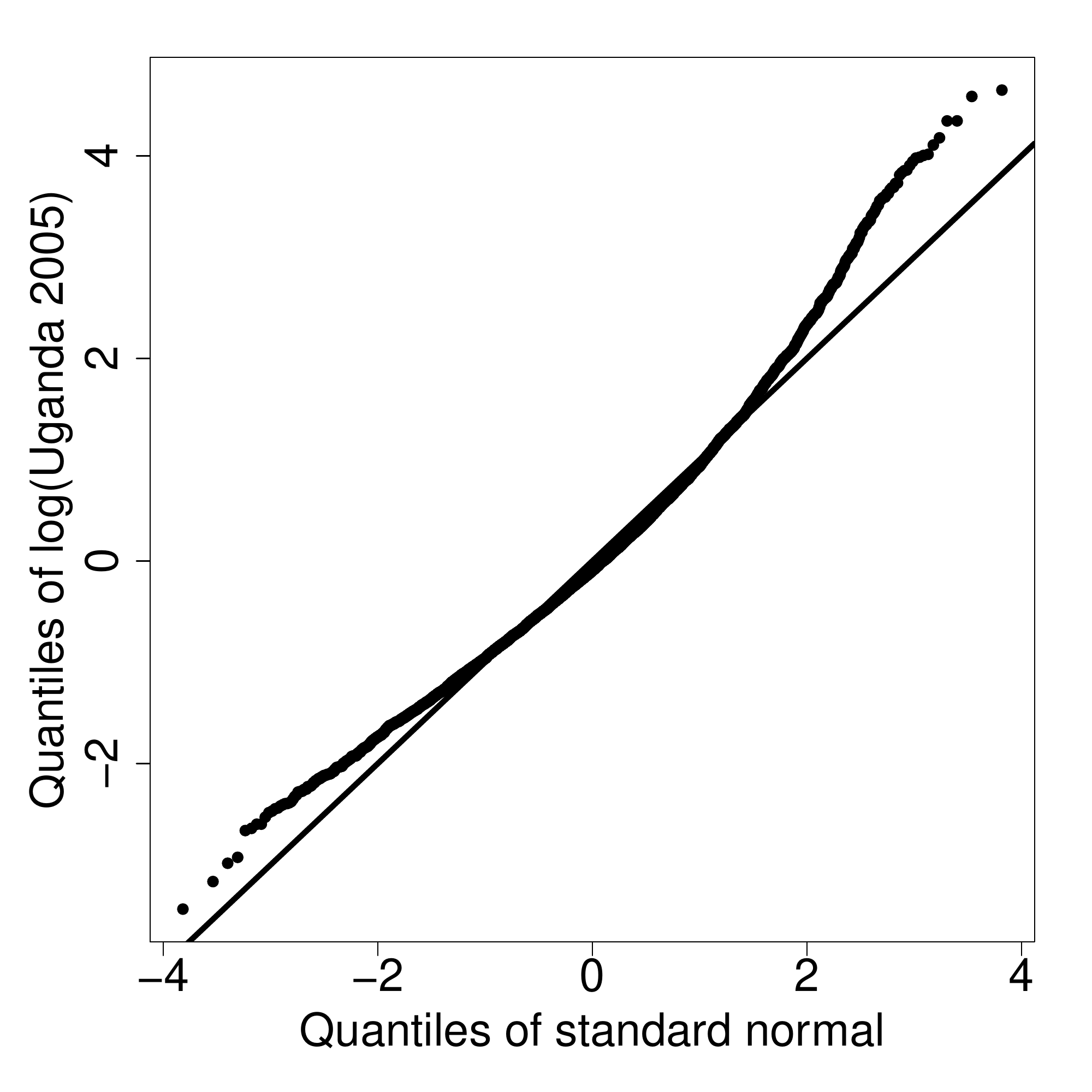}
\includegraphics[scale=0.255]{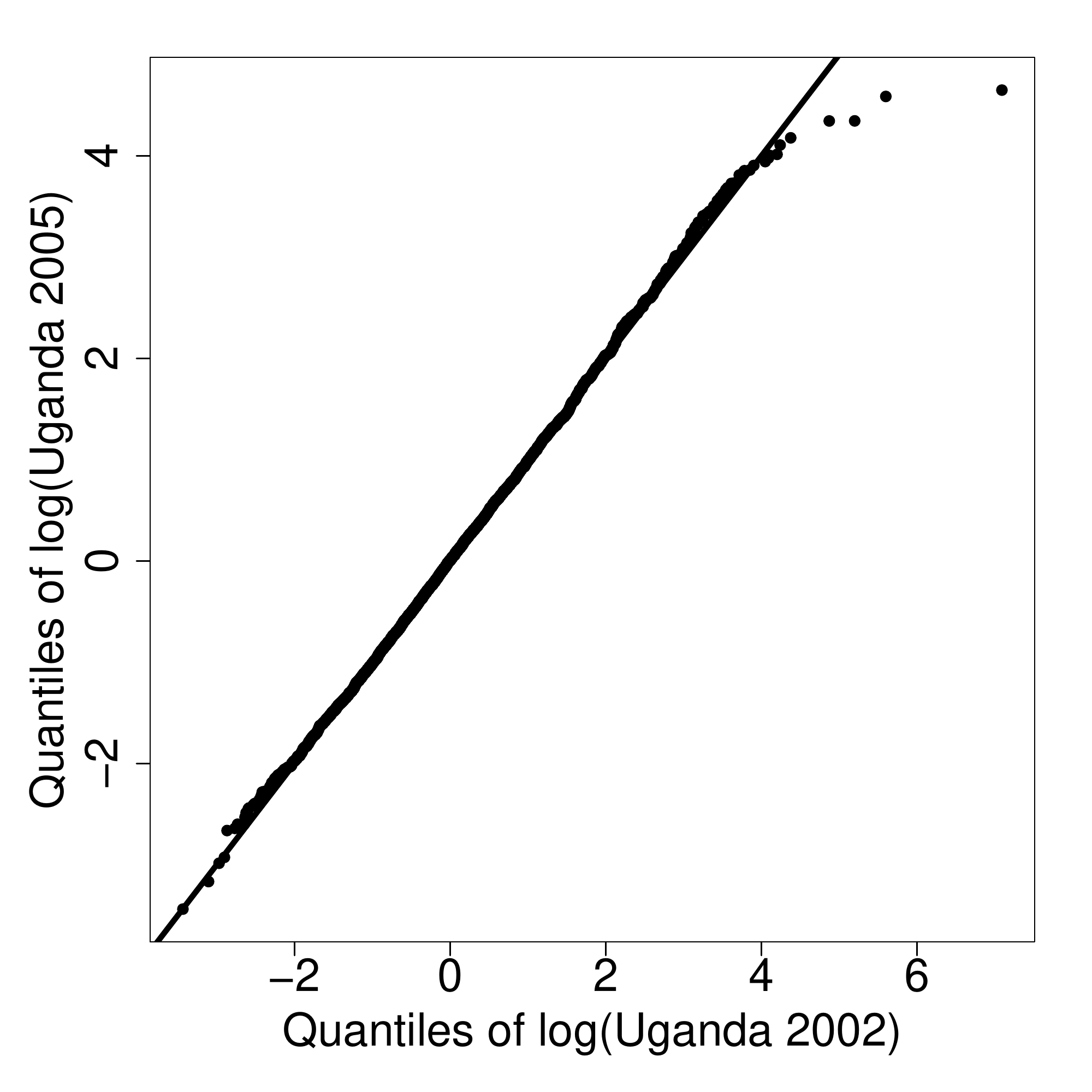}
\caption{QQ-plots of standard normal quantiles against standardised log-transformed Uganda expenditure data for 2002 (left) and  2005 (middle), as well as QQ-plot of standardised log-transformed Uganda expenditure data for 2002 against 2005 (right).}
\label{fig:qqplots1}
\end{figure}

First, we consider the growth incidence curve for the time from $2002$ to $2005$. Inspecting  in Figure \ref{fig:qqplots1} QQ-plots of the standardised log-transformed data (left and middle), we can deduce that both samples show slight departures from the log-normal distribution, but differ from each other only in location and scale, up to four outliers. Hence, we can estimate $\widehat{s}$ according to (\ref{eq:s}) and set $c_{s} = 0$.

\begin{figure}[h!]
\includegraphics[width=0.49\textwidth]{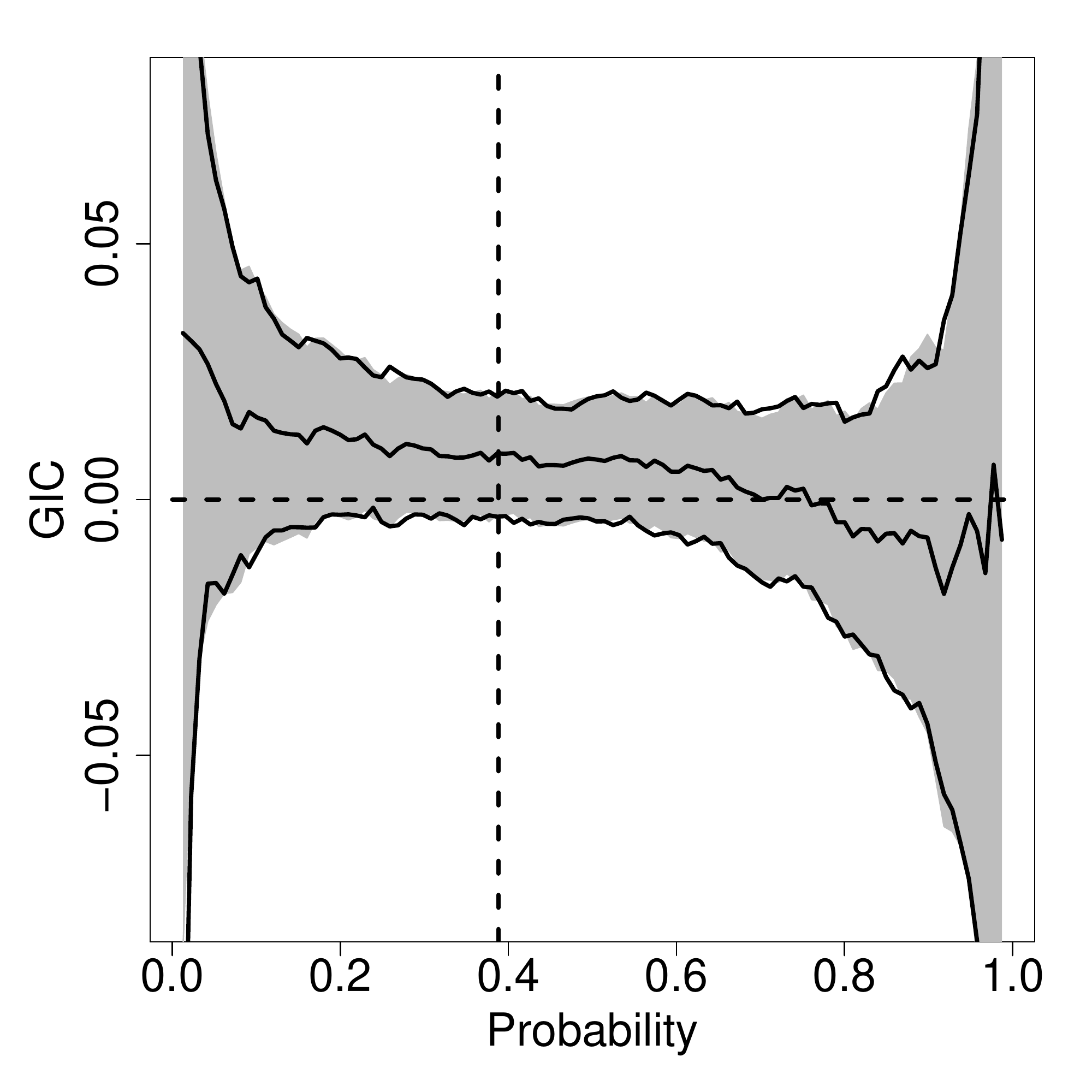}
\quad
\includegraphics[width=0.49\textwidth]{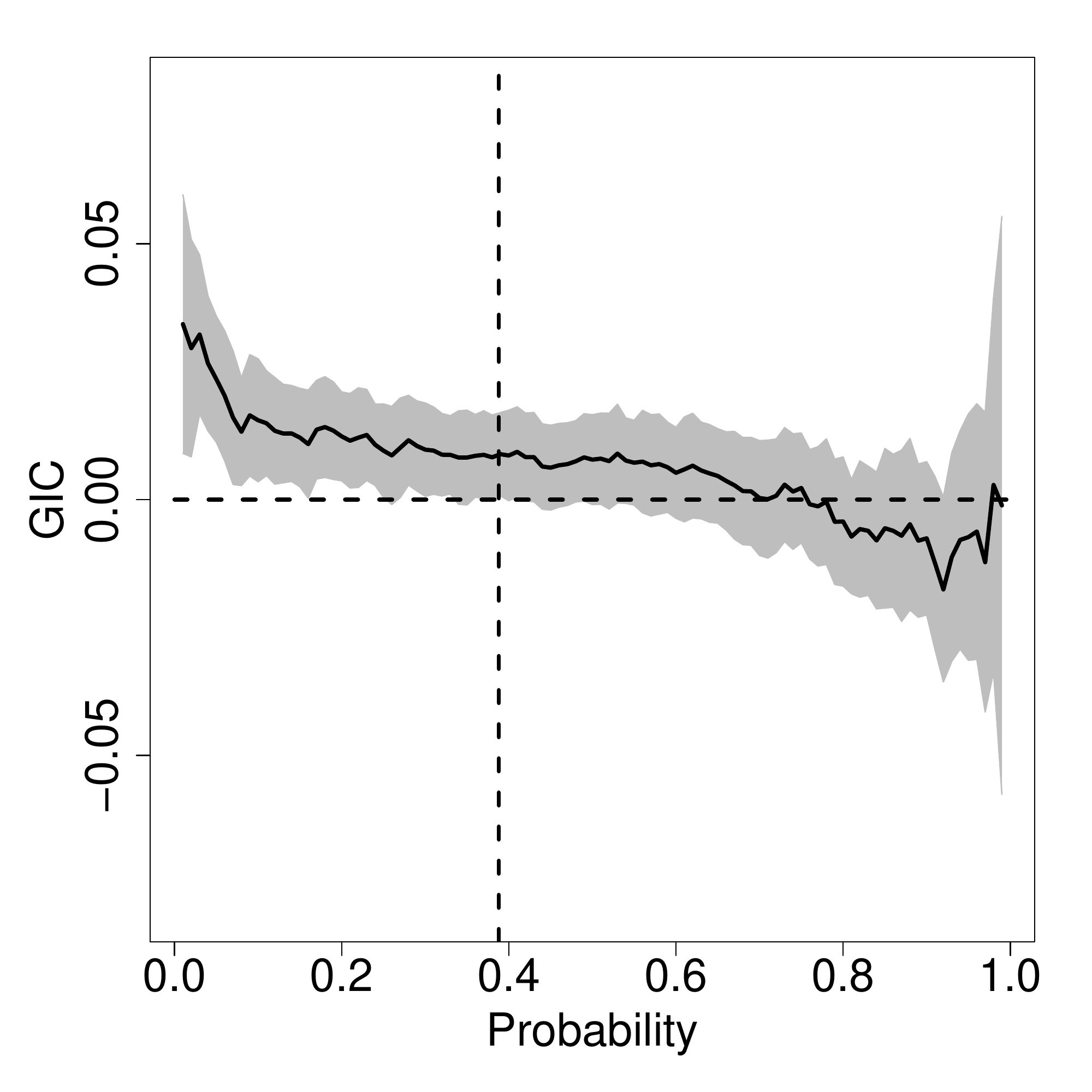} 
\caption{Growth incidence curve for the Uganda data from $2002$ to $2005$ with $95\%$ confidence bands and national poverty line. Simultaneous confidence bands are shown in the left plot, while pointwise confidence bands with the World Bank algorithm in the right plot. }\label{fig:uganda2}
\end{figure}

The estimated growth incidence curve shown in Figure \ref{fig:uganda2} is close to $0$ on the whole interval $(0,1)$. It takes positive values up to the $0.7$ quantile and negative values for higher incomes. The slope tends to be negative. This might suggest that absolute poverty and relative poverty was reduced, and growth was pro-poor according to the weak absolute and relative definition. Both simultaneous confidence bands are shown in the left panel; the grey area corresponds to the plug-in confidence bands, while bold lines are the direct confidence bands. As in simulations for large samples, both approaches lead to nearly the same bands. Simultaneous confidence bands include the zero line, which suggests that none of the discussed effects is in fact significant. In contrast, the considerably tighter confidence bands of the World Bank Toolkit, shown in the right plot, would wrongly suggest otherwise, over-interpreting the non-significant poverty reduction and pro-poor growth.

\begin{figure}[h!]
\includegraphics[scale=0.255]{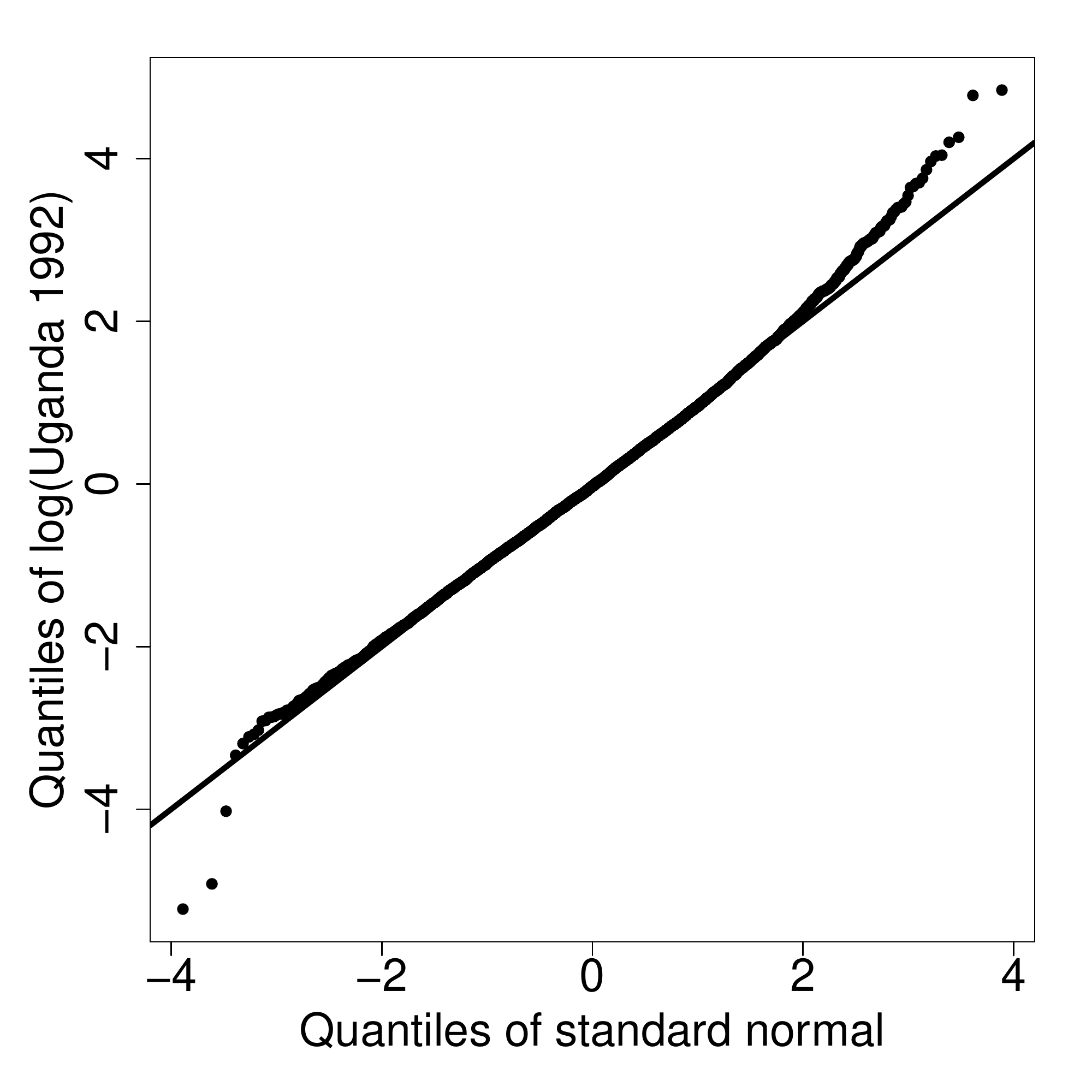}
\includegraphics[scale=0.255]{UgandaQQ02}
\includegraphics[scale=0.255]{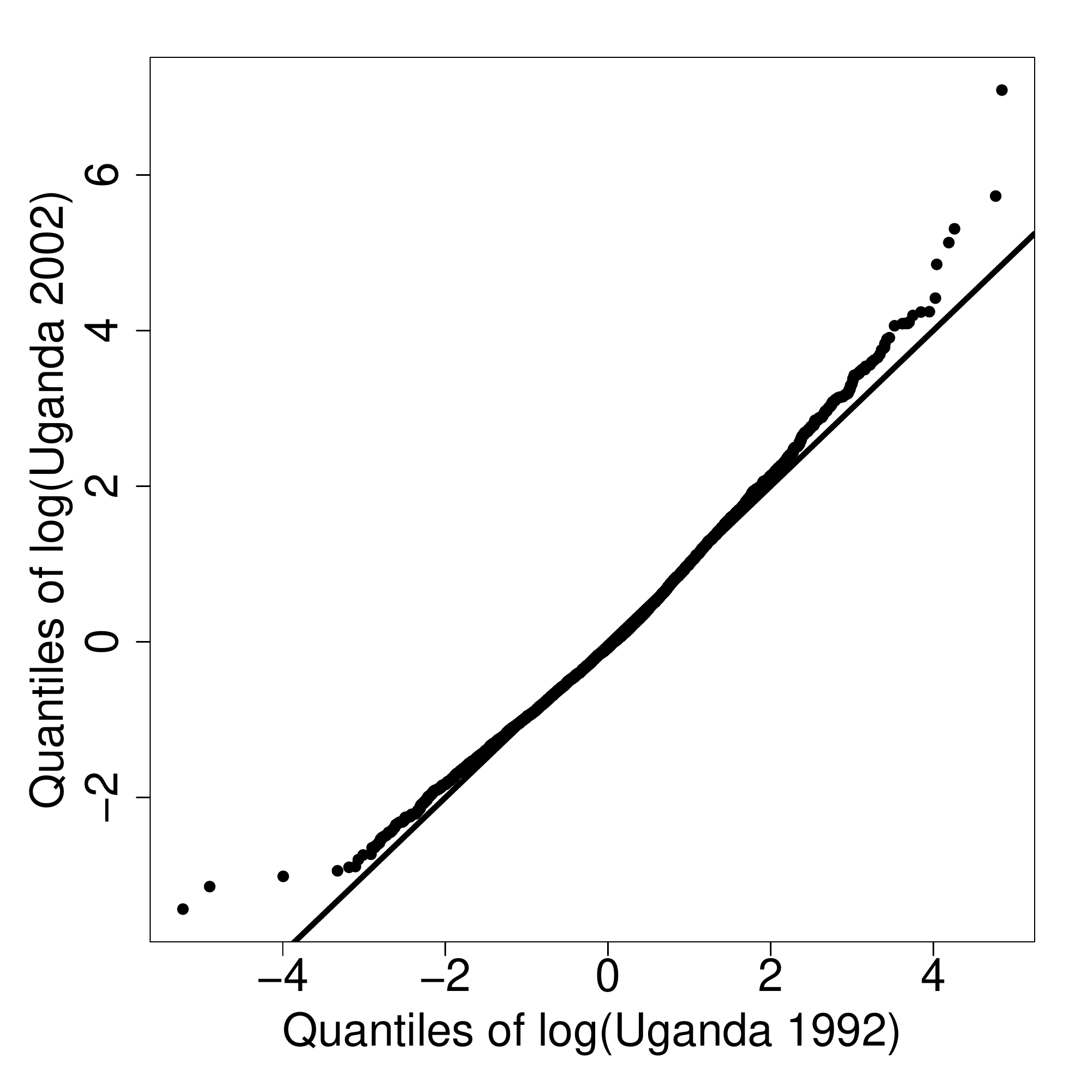}
\caption{QQ-plots of standard normal quantiles against standardised log-transformed Uganda expenditure data for 1992 (left) and  2002 (middle), as well as QQ-plot of standardised log-transformed Uganda expenditure data for 1992 against 2002 (right).}
\label{fig:qqplots2}
\end{figure}

Let us now consider the expenditure data from $1992$ and $2002$. Inspecting QQ-plots of standardised log-transformed data shown in Figure \ref{fig:qqplots2} we find that both data sets are not log-normal and distributions of both data sets differ from each other not only in location and scale. Hence, for the plug-in confidence bands correction $c_s$ needs to be estimated.

\begin{figure}[h!]
\includegraphics[width=0.49\textwidth]{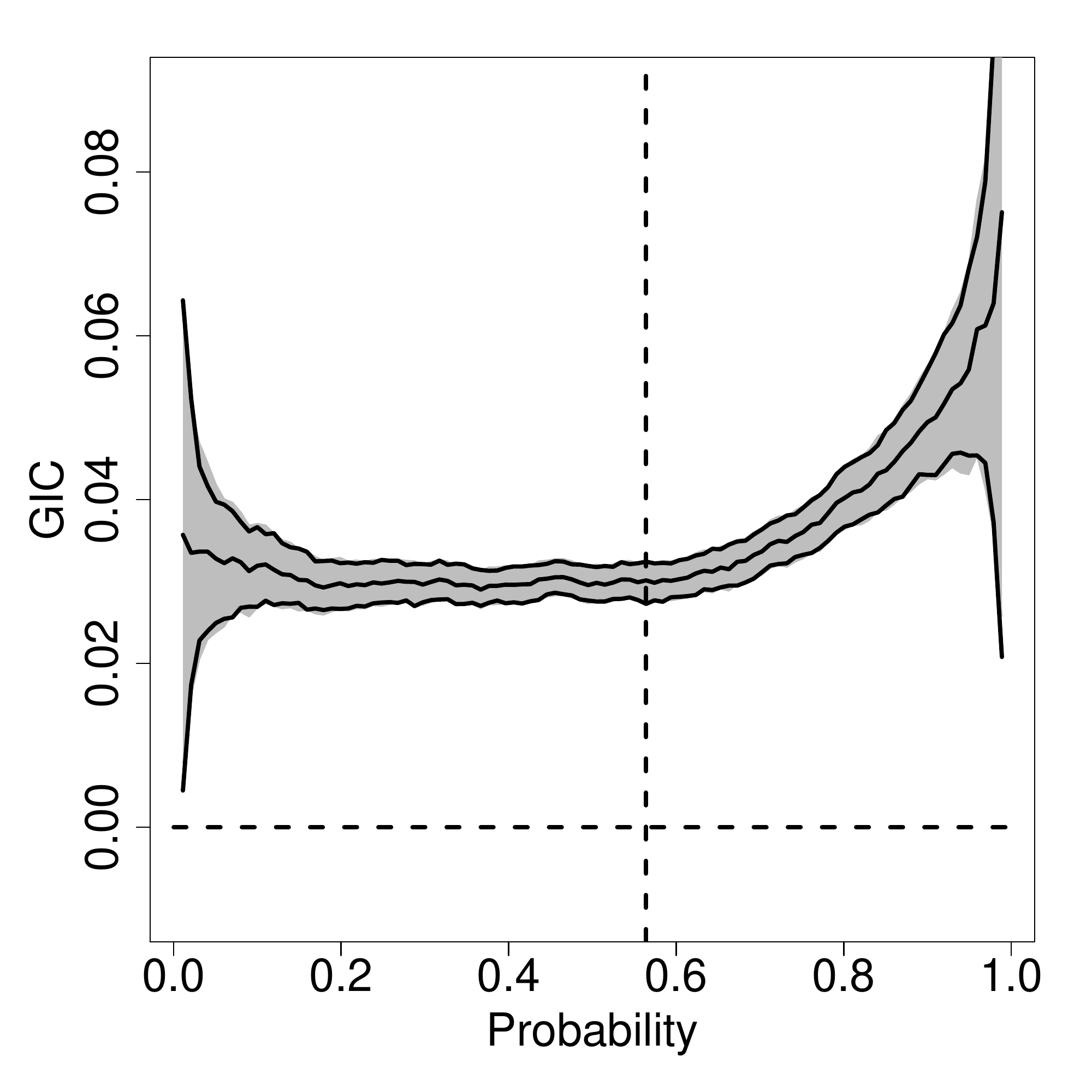}\quad
\includegraphics[width=0.49\textwidth]{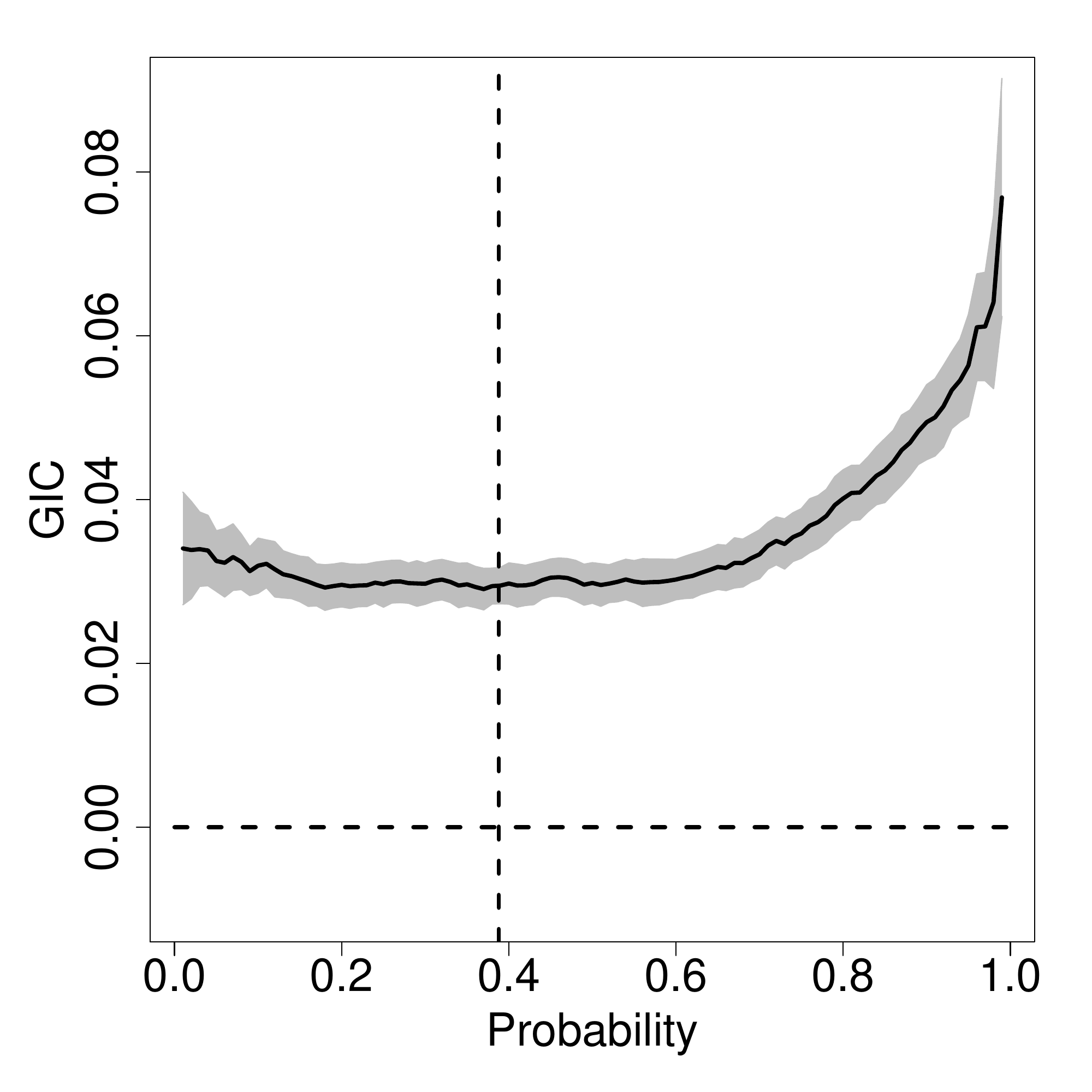} 
\caption{Growth incidence curve for the Uganda data from $1992$ to $2002$ with $95\%$ confidence bands and national poverty line. Simultaneous confidence bands are shown in the left plot, while pointwise confidence bands with the World Bank algorithm in the right plot.}\label{fig:uganda1}
\end{figure}

Figure \ref{fig:uganda1} shows annualized growth incidence curves for Uganda form $1992$ to $2002$ together with the simultaneous confidence band (left) and with the World Bank Toolkit confidence band (right). The estimated growth incidence curve is positive for all quantiles and simultaneous confidence band does not include the zero line.  Absolute poverty was reduced between these two periods, and growth was pro-por using the weak absolute definition. In addition, the growth incidence curve seems to have no significant slope for the poor and a slightly positive slope for the population above the poverty line. This suggests that inequality among the non-poor increased. The confidence band gives evidence that the overall slope of the growth incidence curve on the interval $[0.6,1)$ was non-negative. Confidence bands of the World Bank Toolkit do not allow for such inference about the slope by definition.

\section{Conclusion}\label{sec:conclusion}
Motivated by the concept of growth incidence curves introduced in poverty research we considered the ratio of quantile functions as a tool to compare two distributions. 
We have developed an analytical method for calculating simultaneous confidence bands for ratios of quantile functions and growth incidence curves. Our method requires no re-sampling techniques and rather relies on the asymptotic distribution of the difference of two quantile functions and therefore readily provides simultaneous confidence bands also for the quantile treatment effect, considered as a curve. In the application to the expenditure data from Uganda we demonstrated how simultaneous confidence bands can be used for inference about growth incidence curves and showed that these simultaneous confidence bands are more appropriate than those provided by the World Bank Toolkit.

\section*{Acknowledgments} The authors acknowledge support by the Ministry of Education and Cultural Affairs of Lower Saxony in the project Reducing Poverty Risk. We also thank Gordon Sch\"ucker for translating the World Bank algorithm for GIC confidence intervals from Stata to R.

\bibliography{gic}
\bibliographystyle{apalike}

\appendix

\section{Appendix}

\subsection{Proofs of Section \ref{sec:estimation}}
To prove Theorem \ref{the:ass_lognorm} and Corollary \ref{cor:ass_norm} we use the following standard result.

\begin{theorem}[\citealp{Cramer:46}, p. 368--369]\label{the:quant_dist}
Let $X$ be a random variable with cumulative distribution function $F$, which is continuously differentiable at some $x$ with $F(x) = p$ and $F^{\;'}(x)>0$. Let also $Q(p)=F^{-1}(p)$ denote the quantile function, $q(p)=Q^{'}(p)=1/F^{\;'}\{Q(p)\}$ the quantile density and $\widehat{Q}(p)$ the sample quantile function.
\begin{itemize}

\item[(i)] The distribution of $\widehat{Q}(p)$ is asymptotically normal with mean $Q(p)$ and variance 
$
n^{-1}p(1-p)\{q(p)\}^2
$
for $n\rightarrow\infty$ and for every $p \in (0,1)$.

\item[(ii)] If in addition $F$ is continuously differentiable at some $\tilde x$ with $F(\tilde x) = \tilde p$ and $F^{'}(\tilde{x})>0$ for $p \le \tilde p$, then the joint distribution of $\{\widehat{Q}(p), \widehat{Q}(\tilde p)\}$ is asymptotically bivariate normal with expectation $\{Q(p),Q(\tilde{p})\}$ and 
$
\mathop{Cov}\{Q(p), Q(\tilde p)\} = n^{-1}p(1-\tilde p)q(p)q(\tilde p)
$
for $n\rightarrow\infty$ and for every $p \in (0,1)$.
\end{itemize}
\end{theorem}

Theorem \ref{the:ass_lognorm} shows that the distribution of $\{\widehat{g}(p)\}^m$ can be approximated by a log-normal distribution.\\

{\bf Proof of Theorem \ref{the:ass_lognorm}}

From (\ref{eq:log_trafo}) and Theorem \ref{the:quant_dist}, estimator 
$
\log\{\hatG(p)+1\}  =m \log\{g(p)\}=m \{\widehat{\calQ}_{2}(p) - \widehat{\calQ}_{1}(p)\}
$
is the sum of two asymptotically normal estimators. Since $X_1 $ and $ X_2$ are independent, their sum is asymptotically normal with the mean
$$
\mu(p)=m \{\calQ_2(p)-\calQ_1(p)\}=m[\log\{Q_1(p)\}-\log\{Q_2(p)\}]=m\log\{g(p)\}
$$
and variance
$$
\sigma^2(p)= m^2p(1-p)\left[\frac{\{q_1(p)\}^2}{n_1}+\frac{\{q_2(p)\}^2}{n_2}\right].
$$
Hence, $\{\widehat{g}(p)\}^m$ is log-normally distributed with parameters $\mu(p)$ and $\sigma(p)$. This proves part $(i)$ of the theorem. Part $(ii)$ follows in the same way from Theorem \ref{the:quant_dist} $(ii)$.
$\qed$\\

{\bf Proof of Corollary \ref{cor:ass_norm}}

From Theorem \ref{the:ass_lognorm} we have that $\log\{\widehat{G}(p)+1\}$ is asymptotically normal with parameters $\mu(p)$ and $\sigma(p)$. Let 
$$
Y=\frac{\widehat{G}(p)+1-\exp\{\mu(p)\}}{\exp\{\mu(p)\}\sigma(p)}.
$$
Then, the distribution function of $Y$ is given by
\beqn
F(Y\leq y)&=&F\left[\widehat{G}(p)+1\leq  y\exp\{\mu(p)\}\sigma(p)+\exp\{\mu(p)\}\right]\\
&=&F\left(\frac{\log\{\widehat{G}(p)+1\}-\mu(p)}{\sigma(p)}\leq \frac{\log \left[\exp\{\mu(p)\}\{y\sigma(p)+1\}\right]-\mu(p)}{\sigma(p)}\right)\\
&=&\Phi\left[ \frac{\log \left\{y\sigma(p)+1\right\}}{\sigma(p)} \right] +\o(1)=\Phi\left[y-\frac{y^2\sigma(p)}{2}+\o\{\sigma(p)\}\right]+\o(1),
\eeqn
where $\Phi$ is the cumulative distribution function of a standard normal distribution. 
Since $\sigma(p)\rightarrow 0$ as $\min\{n_1,n_2\}\rightarrow\infty$, the results follows.
$\qed$\\

The proof of Theorem \ref{the:gic_process} relies on the following theorem as given in \cite{Csorgo:83}.
\begin{theorem}[Theorem 3.2.4 in \citealp{Csorgo:83}]\label{the:quant_process}
Let $X$ be a random variable with the cumulative distribution function $F(x)$, quantile function $Q(p)$ and quantile density function $Q^{\;'}(p)=1/F^{\;'}\{Q(p)\}$, $p\in(0,1)$. Let $X_1,\ldots,X_n$ be i.i.d. sample of $X$ and $\widehat{Q}(p)$ be the empirical quantile function as given in (\ref{eq:sample_quantile_fct}). 
Under Assumption \ref{ass:diff} with $X=\calX_1=\calX_2$ there exists a Brownian bridge $\{B_n(p); 0\le p \le 1\}$ such that
\[
\sup_{p \in [\delta_n,1-\delta_n]} \left|\frac{\widehat{Q}(p)-Q(p)}{Q^{\;'}(p)/\sqrt{n}} - B_n(p)\right| \stackrel{a.s.}{=} \O\left\{n^{-1/2}\log(n)\right\}
\]
with $\delta_n = 25n^{-1}\log\log(n)$. If in addition Assumption \ref{ass:tail} (i) holds, there exists a Brownian bridge $\{B_n(p); 0\le p \le 1\}$ such that
\[
\sup_{p \in [0,1]} \left|\frac{\widehat{Q}(p)-Q(p)}{Q^{\;'}(p)/\sqrt{n}} - B_n(p)\right|  \stackrel{a.s.}{=} \O\left\{n^{-1/2}\log(n)\right\}.
\]
If Assumptions \ref{ass:diff} and \ref{ass:tail} (ii) hold, there exists a Brownian bridge $\{B_n(y); 0\le y \le 1\}$ such that
\begin{align}
\begin{split}\label{eq:approx_rates}
\sup_{p \in (0,1)} \left|\frac{\widehat{Q}(p)-Q(p)}{Q^{\;'}(p)/\sqrt{n}} - B_n(p)\right| 
&\stackrel{a.s.}{=}
\begin{cases}
\O\left\{n^{-1/2}\log(n)\right\} & \mbox{if } \gamma < 2\\
\O\left[n^{-1/2}\{\log\log(n)\}^\gamma \{\log(n)\}^{(1+\varepsilon)(\gamma-1)}\right] & \mbox{if } \gamma \ge 2
\end{cases}
\end{split}
\end{align}
for arbitrary $\varepsilon > 0$.
\end{theorem}

{\bf Proof of Theorem \ref{the:gic_process}}

According to Theorem \ref{the:quant_process} there exist series of Brownian bridges $B_{n_1}$ and $B_{n_2}$ such that for $j=1,2$
\[
\sup_{p \in [\delta_{n_j},1-\delta_{n_j}]} \left|\frac{\widehat{\calQ}_j(p)-\calQ_j(p)}{q_j(p)/\sqrt{n_j}} - B_{n_j}(p)\right| \stackrel{a.s.}{=} \O\left\{{n_j}^{-1/2}\log({n_j})\right\}.
\]
This entails
$$
\sup_{p \in [\delta_{n_1},1-\delta_{n_1}]} \left|\sqrt{\frac{s^2 n_2}{n_1+s^2 n_2}}\left\{\frac{\widehat{\calQ}_1(p)-\calQ_1(p)}{q_1(p)/\sqrt{n_1}} -B_{n_1}(y)\right\}\right| \stackrel{a.s.}{=} \O\left\{\sqrt{\frac{n_2}{n_1( n_1+n_2)}}\log({n_1})\right\}
$$
and
$$
\sup_{p \in [\delta_{n_2},1-\delta_{n_2}]} \left|\sqrt{\frac{n_1 }{n_1+s^2 n_2}}\; \left\{\frac{\widehat{\calQ}_2(p)-\calQ_2(p)}{q_2(p)/\sqrt{n_2}} - B_{n_2}(p)\right\} \right| \stackrel{a.s.}{=} \O\left\{\sqrt{\frac{n_1}{n_2( n_1+n_2)}}\log({n_2})\right\}.
$$
The triangular inequality implies together with $n=\min\{n_1,n_2\}$
$$
\sup_{p \in [\delta_{n},1-\delta_{n}]} \Bigg|\sqrt{\frac{n_1n_2}{n_1+s^2 n_2}}\left\{s\; \frac{\widehat{\calQ}_1(p)-\calQ_1(p)}{q_1(p)} -\frac{\widehat{\calQ}_2(p)-\calQ_2(p)}{q_2(p)}\right\}
- B_{n_1,n_2}(p)\Bigg| \stackrel{a.s.}{=} \O\left\{n^{-1/2}\log({n})\right\},
$$
where
\[
B_{n_1,n_2}(p)=\sqrt{\frac{s^2 n_2}{n_1+s^2 n_2}}B_{n_1}(p) - \sqrt{\frac{ n_1}{n_1+s^2 n_2}}B_{n_2}(p).
\]
By the independence of $B_1$ and $B_2$ it follows that $B_{n_1,n_2}$ is a Brownian bridge as well. The other parts of the theorem are proved in the same way.
$\qed$\\

{\bf Proof of equation (\ref{eq:s})}\\
Assumption \ref{ass:scale} states that $q_1(p)=s\;q_2(p)$, which is equivalent to
$f_2\{\calQ_2(p)\}=s\;f_1\{\calQ_1(p)\}$. Function $f_j\{\calQ_j(p)\}$ is known as the density quantile function. This function is positive on its support $[0,1]$. However, this is not a valid density function, since it does not integrate to $1$. Indeed, making a variable change $\calQ_j(p)=x$ implies
$$
\alpha_j=\int_0^1f_j\{\calQ_j(p)\}dp=\int_{-\infty}^\infty \{f_j(x)\}^2dx,\;\;j=1,2.
$$
Therefore, $f_2\{\calQ_2(p)\}=s\;f_1\{\calQ_1(p)\}$ if and only if $s=\alpha_2/\alpha_1$. 
$\qed$\\

{\bf Proof of Lemma \ref{lem:itereated_log}}\\
Following the proof of Theorem \ref{the:gic_process}, it is easy to see that 
$$
\sqrt{\frac{n_1 n_2}{n_1 +  s^2n_2}}\;\left\{\frac{\widehat{\calQ}_1(p)-\calQ_1(p)}{q_1(p)/s}+\frac{\widehat{\calQ}_2(p)-\calQ_2(p)}{q_2(p)}
\right\}.
$$
in $D_{n_1,n_2}^*(p;s)-D_{n_1,n_2}(p;s)$
converges uniformly to a Brownian bridge. Applying the law of iterated logarithm for weighted quantile processes (Theorem 1 and Remark 3 in \citealp{EM:88}) with weight function $[p(1-p)]^\nu$ yields the lemma.
$\qed$

\subsection{Proofs of Section \ref{sec:conf_bands}}

{\bf Proof of Theorem \ref{the:gic_conf1}}

The result follows from the Consequence 4.1.2 on p. 34 of \citet{Csorgo:83}, Theorem \ref{the:gic_process} and Lemma \ref{lem:itereated_log}.$\qed$\\

{\bf Proof of Theorem \ref{the:gic_conf2}}

From Corollary 1 in \citep{CR:84} we can get under Assumptions \ref{ass:basic} and \ref{ass:diff} that
$$
\sup_{p\in[\varepsilon_n,1-\varepsilon_n]}\left|\widehat\calQ_j\left(p+\frac{c_\alpha}{\sqrt{n_j}}\right)-\calQ_j(p)-c_\alpha-B_{n_j}(p)\right|\overset{a.s.}{=}\o_p(1)
$$
and 
$$
\sup_{p\in[\varepsilon_n,1-\varepsilon_n]}\left|\widehat\calQ_j\left(p-\frac{c_\alpha}{\sqrt{n_j}}\right)-\calQ_j(p)+c_\alpha-B_{n_j}(p)\right|\overset{a.s.}{=}\o_p(1)
$$
for $j=1,2$, $\varepsilon_n=n^{\delta-1/2}$, $\delta\in(0,1/2)$. With this,
\begin{align*}
\lim_{n_1,n_2 \rightarrow \infty} \Prob\Bigg\{\widehat{\calQ}_2&\left(p-\frac{c_\alpha}{\sqrt{2n_2}}\right)- \widehat{\calQ}_1\left(p+\frac{c_\alpha}{\sqrt{2n_1}}\right) \leq\calQ_2(p)-\calQ_1(p)\\
&\leq \widehat{\calQ}_2\left(p+\frac{c_\alpha}{\sqrt{2n_2}}\right)- \widehat{\calQ}_1\left(p-\frac{c_\alpha}{\sqrt{2n_1}}\right);\;\varepsilon_n \le p \le 1-\varepsilon_n\Bigg\}\\
&= P\left\{\sup_{p\in[0,1]}|B_{1,n_1}(p) + B_{2,n_2}(p)|\le \sqrt{2}c_\alpha\right\}.
\end{align*}
From the independence on Brownian bridges for $j=1$ and $j=2$ follows
\beqn
P\left\{\sup_{p\in[0,1]}\left|B_{1,n_1}(p) + B_{2,n_2}(p)\right|\le \sqrt{2}c_\alpha\right\} &=& P\left\{\sup_{p\in[0,1]}\left|\sqrt{2}B(p)\right|\le \sqrt{2}c_\alpha\right\} \\
&=& P\left\{\sup_{p\in[0,1]}\left|B(p)\right|\le c_\alpha\right\}
\eeqn
for some Brownian bridge $B$.
$\qed$

\end{document}